\documentclass[12pt, a4, preprint]{aastex}
\usepackage{graphicx}
\begin{document}

\title{The radial distribution of water ice and chromophores across Saturn's system}

\author{Filacchione, G.\altaffilmark{1}, Capaccioni, F.\altaffilmark{2}, Clark, R.N.\altaffilmark{3}, Nicholson, P.D.\altaffilmark{4}, 
Cruikshank, D.P.\altaffilmark{5}, Cuzzi, J.N.\altaffilmark{5}, Lunine, J.I.\altaffilmark{4}, Brown, R.H.\altaffilmark{6}, Cerroni, P.\altaffilmark{2}, Tosi, F.\altaffilmark{2}, Ciarniello, M.\altaffilmark{2}, Buratti, B.J.\altaffilmark{7}, Hedman, M.M.\altaffilmark{4}, Flamini, E.\altaffilmark{8}}

\altaffiltext{1}{INAF-IAPS, Istituto di Astrofisica e Planetologia Spaziali, Area di Ricerca di Tor Vergata, via del Fosso del Cavaliere, 100, 00133, Rome, Italy. Email: gianrico.filacchione@iaps.inaf.it}
\altaffiltext{2}{INAF-IAPS, Istituto di Astrofisica e Planetologia Spaziali, Area di Ricerca di Tor Vergata, via del Fosso del Cavaliere, 100, 00133, Rome, Italy.}
\altaffiltext{3}{US Geological Survey, Federal Center, Denver, CO, 80228, USA}
\altaffiltext{4}{Cornell University, Astronomy Department, 418 Space Sciences Building, Ithaca, NY 14853, USA}
\altaffiltext{5}{NASA Ames Research Center, Moffett Field, CA 94035-1000, USA}
\altaffiltext{6}{Lunar Planetary Laboratory, University of Arizona, Kuiper Space Sciences 431A, Tucson, AZ, USA}
\altaffiltext{7}{NASA Jet Propulsion Laboratory, California Institute of Technology, 4800 Oak Grove Drive, Pasadena, CA 91109, USA}
\altaffiltext{8}{ASI, Italian Space Agency, viale Liegi 26, 00198, Rome, Italy}

\begin{abstract}
Over the last eight years, the Visual and Infrared Mapping Spectrometer (VIMS) aboard the Cassini orbiter has returned hyperspectral images in the 0.35-5.1 $\mu m$ range of the icy satellites and rings of Saturn. These very different objects show significant variations in surface composition, roughness and regolith grain size as a result of their evolutionary histories, endogenic processes and interactions with exogenic particles. The distributions of surface water ice and chromophores, i.e. organic and non-icy materials, across the saturnian system, are traced using specific spectral indicators (spectral slopes and absorption band depths) obtained from rings mosaics and disk-integrated satellites observations by VIMS. Moving from the inner C ring to Iapetus, we found a marking uniformity in the distribution of abundance of water ice. On the other hand the distribution of chromophores is much more concentrated in the rings particles and on the outermost satellites (Rhea, Hyperion and Iapetus). A reduction of red material is observed on the satellitesÕ surfaces orbiting within the E ring environment probably due to fine particles from Enceladus' plumes. Once the exogenous dark material covering the IapetusÕ leading hemisphere is removed, the texture of the water ice-rich surfaces, inferred through the 2 $\mu m$ band depth, appears remarkably uniform across the entire system.
\end{abstract}
\maketitle

\section{Introduction}
Saturn, the sixth planet orbiting the Sun at an average distance of 9.56 AU, has one of the most complex system of satellites and ring in our solar system. The composition of these objects is the result of a particular distribution of primordial material inside the proto planetary disk \cite{bib1} and in the circum-planetary nebula \cite{bib2}, and of the subsequent evolution caused by large impacts, meteoroid bombardment, weathering and interaction with magnetospheric and exogenic particles. The satellites' surfaces are then shaped by endogenic processes like cryovolcanism, tectonism and resurfacing processes \cite{bib3}. 

The entire Saturn system was formed in the outer part of the proto-planetary disk, in a region well beyond the Òsnow-lineÓ, currently placed at a distance of about 2.7 AU from the Sun, where water and volatiles condense in ices \cite{bib4}. The study of Saturn's current ring and satellite composition is fundamental to constrain different formation scenarios and evolutionary models \cite{bib1}. For this reason we have investigated how water ice and chromophores, e.g. organic and non-ice materials, are distributed on both ring and satellite surfaces by using observations in the 0.35-5.1 $\mu m$ spectral range returned by VIMS (Visual and Infrared Mapping Spectrometer) \cite{bib5} aboard the Cassini mission. 

The variability of water ice and chromophores across the icy surfaces of the SaturnÕs system is traced using specific spectral indicators applied to a high-quality and extensive sub-set of VIMS data, consisting of 2264 disk-integrated observations of the icy satellites and several ring mosaics collected from 2004 until June 2010 \cite{bib6}. We have adopted three spectral indicators that represent the best tracers for presence of chromophores and water ice: the spectral slopes in the 0.35-0.55 and 0.55-0.95 $\mu m$ ranges and the 2 $\mu m$ water ice band depth \cite{bib7, bib8}. Chromophores-rich surfaces appear in general more red in the visible spectral range and consequently have positive slopes. Water ice-rich surfaces are characterized by blue or neutral slopes and intense 2.0 $\mu$m band depth. As discussed in \cite{bib6}, the water ice band depth is related to the regolith grain size: for a fixed phase angle, small grains (10 $\mu m$) have high reflectance at continuum level but small band depth. On contrary, large grains (100 $\mu m$) are characterized by a lower reflectance and larger band depth.

VIMS spectral slopes, expressed in $\mu m^{-1}$, are measured through a best linear fit to the reflectance spectra in the 0.35-0.55 and 0.55-0.95 $\mu m$ ranges. Before the fit, the reflectance is normalized to 1 at 0.55 $\mu m$ to remove illumination effects and separate color variations from brightness \cite{bib9}.

Since both band depth and spectral slopes increase with solar phase, as explained by \citep{bib10} and verified on VIMS data by \cite{bib6}, we have minimized this effect by selecting only observations acquired at intermediate solar phase angles between 20$^\circ$ and 40$^\circ$; within this interval the variability of the spectral parameters due to photometry is minimal allowing us to perform a comparative analysis of the entire population of the rings and satellites under the same illumination conditions.   

\section{Chromophores distribution}

The 0.35-0.55 $\mu m$ spectral slope is sensitive to the effects of intraparticle mixing (e.g. grains within grains) of the chromophores in water ice, while the 0.55-0.95 $\mu m$ slope is an intimate (e.g. salt and pepper mixture) and areal mixing marker \cite{bib11}. Pure water ice is characterized by an almost ÒneutralÓ spectral behavior, resulting in high albedo, a slightly positive 0.35-0.55 $\mu m$ spectral slope and a small negative 0.55-0.95 $\mu m$ spectral slope, with some variations caused by the different effective grain sizes in the regolith \cite{bib6}. Adding chromophores causes the two slopes to increase at a rate that depends upon the contaminant's composition and fractional amount. Meanwhile, the 2 $\mu m$ band depth depends on the relative abundances of ice and dark material, as well as the regolith grain size.

\begin{figure}[h!]
\centering
\includegraphics[height=160mm]{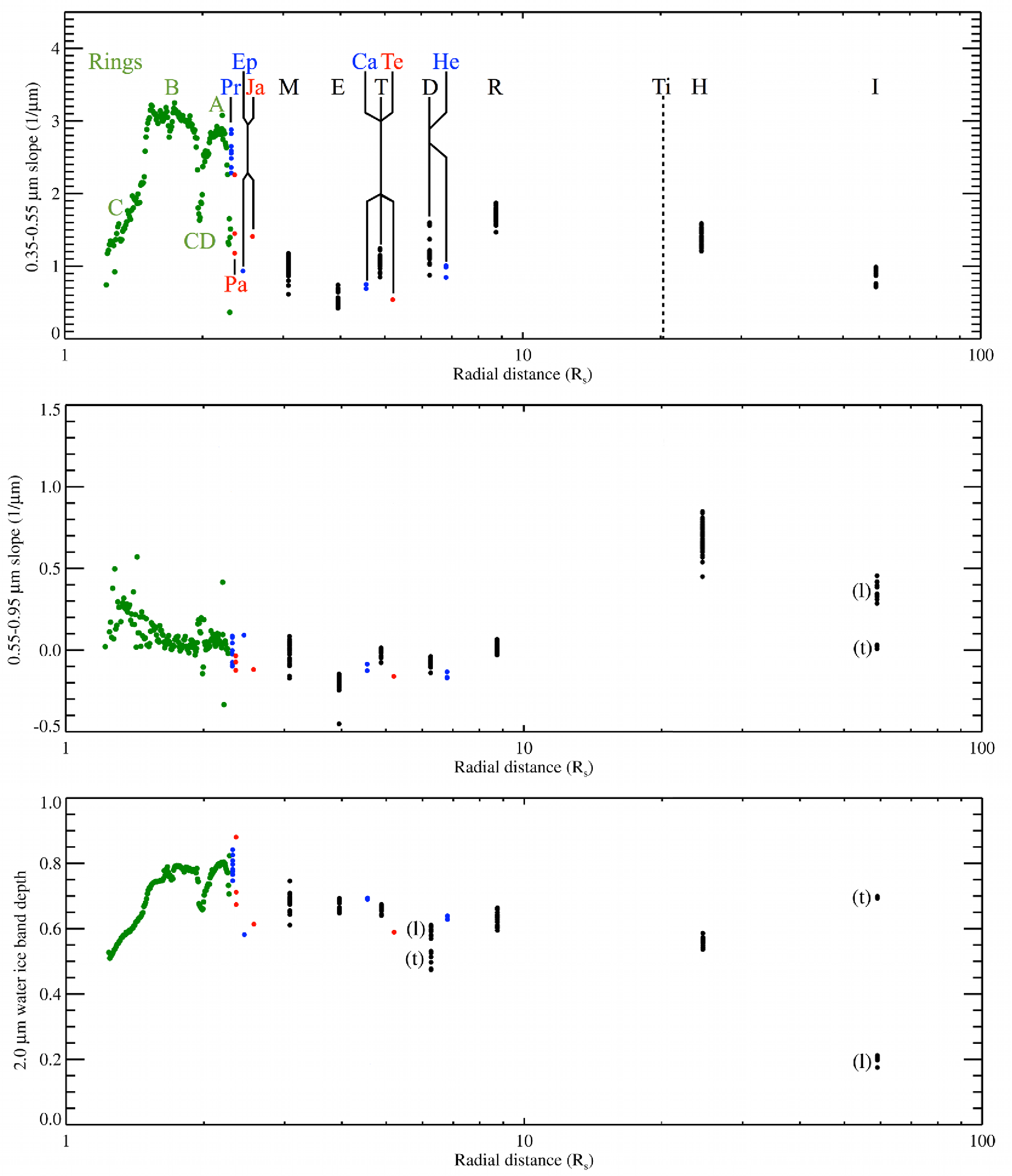}
\caption{Radial profiles of spectral indicators for SaturnÕs ring, minor and regular satellites: visible spectral slope 0.35-0.55 $\mu m$ (top panel), 0.55-0.95 $\mu m$ slope (center panel), water ice 2 $\mu m$ band depth (bottom panel). The ring radial profiles span from inner C ring (73.500 km) to outer A ring (141.375 km) at 400 km/sample resolution. Co-orbiting satellites radial distances are shown with an offset to improve visualization. Spectral quantities are computed from observations taken with solar phases ranging between 20$^\circ$ and 40$^\circ$ to minimize photometric effects on the radial trends. Legend: Prometheus (Pr), Pandora (Pa), Epimetheus (Ep), Janus (Ja), Mimas (M), Enceladus (E), Tethys (T), Calypso (Ca), Telesto (Te), Dione (D), Helene (He), Rhea (R), Titan (Ti), Hyperion (H), Iapetus (I). Leading and trailing hemisphere observations for Dione and Iapetus are indicated with (l) and (t), respectively.}
\label{fig:figure1}
\end{figure}

As shown in Fig. \ref{fig:figure1}, top panel, across the rings we measure an increase of the 0.35-0.55 $\mu m$ spectral slope from inner C ring (between 1.24-1.52 $R_S$; 1 $R_S$, or Saturn's radius, corresponding to 60.268 km) to B ring (1.52-1.95 $R_S$), where the maximum reddening on the Saturnian system is measured (between 3 and 4 $\mu m^{-1}$). In the Cassini division (1.95-2.02 $R_S$) the slope decreases to reach another local maximum at the A ring (2.02-2.27 $R_S$). 

\placefigure{figure1}

Prometheus and Pandora, the two shepherd moon of the F ring, appear different in color: Prometheus in fact is redder than Pandora. Moreover Prometheus is the only moon of the population that shows a reddening similar to A and B rings particles and for this reason it could be the Òmissing linkÓ, able to connect together the origins, evolution and composition of rings and satellites: this result strengthens the hypothesis that the minor satellites were accreted starting from ring material during the viscous spreading of the young, massive disk from which the rings are thought to have formed \cite{bib12, bib13}.  

Moving outwards, the slope is influenced by the presence of the E ring particles, released by the plumes emanating from the Òtiger stripesÓ fractures at the south pole of Enceladus \cite{bib14}: starting from Janus and Epimetheus and out to Mimas, the abundance of chromophores decreases, reaching a minimum on Enceladus. Enceladus shows the bluest spectrum, a distinctive property caused by the presence of very fresh water ice with a negligible amounts of contaminants \cite{bib15}. Continuing to the other satellites orbiting within the outermost part of the E ring environment, from Tethys to Dione and Rhea, we observe a continuous increase of the reddening with the radial distance. However, both Calypso and Telesto, the two Tethys lagrangian satellites, and Helene, one of the two Lagrangians of Dione, appear bluer than Tethys and Dione themselves and similar to the color of Enceladus, indicating that the surfaces of these three small moons are fresher and probably coated by a significant amount of E ring particles. VIMS spectral results are supported by high-resolution images returned by Cassini-ISS in which the surfaces of these moons appear unusually smooth and lacking of small craters, compared to other Saturnian moons \cite{bib13}.  

Regarding the inner regular satellites, the 0.35-0.55 $\mu m$ slope has a bowl-shaped distribution with a minimum on Enceladus and maxima on A-B rings in the inner part and on Rhea in the outer part. RheaÕs reddening, equal to about 1.5 $\mu m^{-1}$, is similar to C ring and Cassini division particles. Beyond Titan (not considered in this analysis since the dense atmosphere does not allow us to accurately retrieve spectral indicators of the surface) we observe a linear decrease of the 0.35-0.55 $\mu m$ slope moving from Hyperion to Iapetus. The decreasing trend continues at Phoebe, the outermost satellite orbiting at 215 $R_S$, for which we have measured a value of about 0.3 $\mu m^{-1}$ at about 90$^\circ$ solar phase: these data have been not used in this work because they are outside the 20$^\circ$-40$^\circ$ range.   

The radial trend for the 0.55-0.95 $\mu m$ slope is shown in Fig. \ref{fig:figure1}, central panel. Across the ring, the distribution shows two local maxima corresponding to the C ring and Cassini division where the slope reaches about 0.3 and 0.2 $\mu m^{-1}$, respectively. A and B rings appear more neutral, with slope running between 0.0 and 0.1 $\mu m^{-1}$. The large difference in 0.35-0.55 $\mu m$ reddening previously discussed between Prometheus and Pandora, is reduced substantially in the 0.55-0.95 $\mu m$ range, with a slightly higher reddening observed on Prometheus. Also in this case we observe a bowl-shaped distribution across the E ring environment, with a minimum corresponding to the orbit of Enceladus. Enceladus, like the three Lagrangian moons Calypso, Telesto and Helene, has a bluer spectrum resulting in the more negative 0.55-0.95 $\mu m$ spectral slope values. Neutral spectral slopes characterize the remaining inner regular satellites. In this spectral range the maximum reddening occurs on the outer satellites: on Hyperion on average is measured a spectral slope of 0.7 $\mu m^{-1}$ between 0.55-0.95 $\mu m$ while on Iapetus 0.4 and about 0.0 $\mu m^{-1}$ on the dark leading and bright trailing hemispheres, respectively. Such high reddening values are the result of the dominant presence of dark materials such as organics (PAH), iron-silicate nanophases or carbon within water ice \cite{bib16, bib17, bib18}. PhoebeÕs 0.55-0.95 $\mu m$ spectral slope, not included in Fig. 1, is equal to about -0.1 $\mu m^{-1}$ at 90$^\circ$ solar phase.

\section{Water ice distribution}
The 2 $\mu m$ water ice band depth (Fig. \ref{fig:figure1}, bottom panel) increases almost linearly between 0.5 to 0.7 moving from inner to outer C ring. B ring shows a dual distribution of the band depth: the inner part of the B ring, for radial distance $< 1.66 \ R_S $, or $<$100.000 km, is below 0.75 while the outer part has higher values, up to 0.8. At the spatial resolution of this dataset (400 km/bin) the ripple between 1.66-1.74 $R_S$, or 100.000-105.000 km, is evident where the band-depth variations are correlated with sharp swings in optical depth, which likely alter the collisional environment experienced by the ring particles. The collisional processes occurring in this region are responsible for water ice resurfacing, resulting in an increase in the measured band depth. The band depth drops to about 0.65 in the middle of the Cassini division and then increases to values similar to the outer B ring across the A ring. Prometheus and Pandora have in average the highest band depth ($>$0.8) among the satellites and as seen previously for the two spectral slopes, also in this case they appear very similar to A and B ring particles. With a band depth of about 0.6, Janus and Epimetheus instead, are more similar to C ring and Cassini division particles. Moving to the regular satellites, from Mimas to IapetusÕ bright trailing hemisphere, we find that the band depth distribution is basically constant around a value of 0.7 with some small negative deviations associated with Dione and Hyperion. It is interesting to note the 2 $\mu m$ band depth differences measured in the Lagrangian systems: a difference of about 0.1 is seen between Calypso (higher) and Telesto (lower) with Tethys lying in between. Similarly, Helene has a stronger band depth than Dione. This could be the consequence of the layering of fresh material coming from E ring environment, which causes differences in surface water ice composition and in regolith grain size between the lagragian moons: higher band depth can be explained with more pure water ice or with larger grains.

On average Dione's band depth is the lowest of the inner regular satellites. This satellite shows a remarkable difference in the distribution of the band depth which appear stronger on the leading hemisphere (indicated by l in fig 1, bottom panel) while on the trailing side (indicated by t), where wispy terrains are located, the band depth is less intense. Another remarkable difference is seen between IapetusÕ bright trailing hemisphere observations, where the 2 $\mu m$ band depth is approximately 0.7 and dark leading hemisphere where it drops below 0.2. On Phoebe, not shown in Fig. 1, we have measured a 2 $\mu m$ water ice band depth of about 0.15-0.2 at 90$^\circ$ solar phase.

\section{Findings}
VIMS data reveal striking differences among ring regions and satellites, ranging from Enceladus and CalypsoÕs bluish surfaces, which appear very bright, water ice-rich and almost uncontaminated, to the more distant Hyperion, Iapetus and Phoebe where metals, organics and carbon dioxide are mixed within water ice, resulting in lower albedos and redder spectra \cite{bib17, bib29}. In general, ring particles appear to have peculiar properties, being the reddest objects in the SaturnÕs system at visible wavelengths while maintaining sharp and intense water ice bands in the infrared range \cite{bib6, bib19, bib20, bib21}. This spectral behavior is compatible with crystalline water ice polluted by chromophores, e.g. organic material, resulting from the irradiation of simple hydrocarbons \cite{bib22, bib23}, nanophase iron or hematite \cite{bib16, bib24}, tholins in intimate mixing \cite{bib11}, amorphous silicates \cite{bib25}, carbonaceous particles \cite{bib26} or different combinations of these endmembers.

The three radial trends shown in Fig. \ref{fig:figure1} allow us to simultaneously retrieve and trace the distribution and mixing of chromophores within water ice particles across the Saturnian system: comparing the 0.35-0.55 $\mu m$ and 0.55-0.95 $\mu m$ spectral slopes profiles, it appears evident that these two quantities have opposite trends across the ring. The 0.55-0.95 $\mu m$ reddening becomes stronger across ring regions having low optical depth like the C ring and the Cassini division where the 0.35-0.55 $\mu m$ slope reaches the minimum values. These trends can be explained with the presence of a small fraction of inclusions of dark material distributed among the ice particles. In contrast, regions showing stronger band depth, like A and B rings, are characterized by higher reddening in the 0.35-0.55 $\mu m$ range while appearing more neutral in the 0.55-0.95 $\mu m$: such spectral behavior is compatible with the presence of UV-blue absorbing chromophores implanted in the water ice matrix. The ring particle's properties derived from VIMS radial profiles are therefore in agreement with the compositional trend resulting from the Òballistic modelÓ \cite{bib26} in which the dark material, a residuum of meteoritic and cometary bombardment, accumulates in the Cassini division and C ring where the optical depth is lower.

Moving to the satellites, two distinctive zones characterize the 0.35-0.55 $\mu m$ and 0.55-0.95 $\mu m$ spectral slopes radial distributions: the inner satellites, orbiting within the E ring environment, and the outer satellites orbiting beyond Titan, which acts like a "barrier" between the rings and satellites inside of its orbit and the satellites and the giant dust ring of Phoebe \cite{bib27} outside its orbit. The 0.35-0.55 $\mu m$ visible slope of the inner satellites has a bowl-shaped distribution centered on Enceladus. The dominant exogenous process here is the release of particles from the plumes of Enceladus \cite{bib14} which, after feeding the vast E ring region, impact with the embedded satellites. 
The outer satellites, Hyperion, Iapetus (and Phoebe), have distinctive compositional properties, differing from the inner ones with respect to the presence of carbon dioxide \cite{bib8, bib28, bib29} and dark material, e.g., hydrocarbons, iron nanophases, and possibly tholins \cite{bib16, bib17, bib30}. The linear decrease of the reddening observed moving outwards from Hyperion to Phoebe is compatible with the deposition of dust particles coming from Phoebe's ring. Therefore E ring and Phoebe's ring play similar roles in influencing the spectral properties of the moons orbiting within their neighborhoods: while the former spreads fresh and bright water ice particles in the inner Saturnian system, the latter causes a contamination of dark and organics-rich material on the outer moons.

\section{Summary}
In conclusion, the surface composition of rings and ice satellites that we measure today is the result of the original chemistry of the circum-planetary nebula from which they condensed and of the subsequent dynamical evolution (meteoroid and exogeneous particle impacts) and geochemical history (solar and high energy particles irradiation, loss of high volatility ices). While the reddening observed at visible wavelengths changes significantly across the Saturnian system, the strength of the water ice band is remarkably uniform once Phoebe and Iapetus' dark material, possibly originated by Phoebe \cite{bib29}, is removed. The reddening variations are probably caused by secular processes while the water ice distribution seems to be more related to the primordial composition of the circum-planetary nebula. Therefore we can deduce that the chemistry occurring at the time of formation is not completely cancelled by the evolution processes of the ring and satellites. Such results can help us to discriminate among the different formation scenarios and to better understand the processes at the origin of the satellites and rings of the outer planets.

\paragraph*{Acknowledgments}
This research has made use of NASA's Astrophysics Data System and was completed thanks to the financial support of the Italian Space Agency (grant I/015/09/0) and NASA through the Cassini project.

\end{document}